\documentclass[conference,letterpaper]{IEEEtran}
\IEEEoverridecommandlockouts%
\usepackage{cite}
\usepackage[para]{footmisc}
\usepackage{amsmath,amssymb,amsfonts}
\usepackage{algorithmic}
\usepackage{graphicx}
\graphicspath{{./img/}}
\usepackage{textcomp}
\usepackage[table,xcdraw]{xcolor}
\usepackage{diagbox}
\usepackage{colortbl}
\usepackage{todonotes}
\usepackage{multirow}
\usepackage{url}
\usepackage{nohyperref} 
\usepackage[per-mode=symbol,group-separator={,},group-minimum-digits={3},binary-units]{siunitx}
\usepackage{subfig}
\usepackage[nohyperlinks,nolist]{acronym} 
\usepackage[]{algorithm2e}
\usepackage{tikz}
\usepackage{tabularx}
\newcolumntype{Y}{>{\centering\arraybackslash}X}
\newcolumntype{L}{>{\arraybackslash}X}
\newcolumntype{R}{>{\raggedleft\arraybackslash}X}
\newcolumntype{C}[1]{>{\centering\arraybackslash}p{#1}}
\usepackage{booktabs}
\usepackage[absolute,showboxes]{textpos}
\usetikzlibrary{patterns}
\usetikzlibrary{spy}
\usepackage{pgfplots}
\pgfplotsset{compat=newest}
\usepgfplotslibrary{units}
\usepgfplotslibrary{groupplots}
\makeatletter
\pgfplotsset{
 unit code/.code 2 args=
   \begingroup
   \protected@edef\x{\endgroup\si{#2}}\x
}
\makeatother
\usetikzlibrary{matrix}

\definecolor{CoreGray}{HTML}{BFBFBF}
\definecolor{CoreBlack}{HTML}{333333}
\definecolor{CoreBlue}{HTML}{002E7D}
\definecolor{CoreGreen}{HTML}{6AAC8E}
\definecolor{CoreRed}{HTML}{C80000}
\definecolor{CoreYellow}{HTML}{E6AC00}
\definecolor{CoreWhite}{HTML}{FFFFFF}
\definecolor{CoreLightBlue}{HTML}{8BBDEB}
\colorlet{LightCoreGray}{CoreGray!20}
\colorlet{LightCoreBlack}{CoreBlack!20}
\colorlet{LightCoreGreen}{CoreGreen!30}
\colorlet{LightCoreRed}{CoreRed!20}
\colorlet{LightCoreYellow}{CoreYellow!20}
\colorlet{LightCoreWhite}{CoreWhite!20}

\NewDocumentCommand{\codeword}{v}{\texttt{\textcolor{CoreBlack}{#1}}}

\usepackage[absolute,showboxes]{textpos}



\begin{document}

\bstctlcite{my:BSTcontrol}

\title{
    Dynamic Service-Orientation for\\ Software-Defined In-Vehicle Networks
}
\author{\IEEEauthorblockN{Timo H\"ackel, Philipp Meyer, Mehmet Mueller, Jan Schmitt-Solbrig, Franz Korf, and Thomas C. Schmidt}\IEEEauthorblockA{\href{http://www.haw-hamburg.de/ti-i}{\textit{Dept. Computer Science}},
\href{http://www.haw-hamburg.de/ti-i}{\textit{Hamburg University of Applied Sciences}}, Germany \\
\{\href{mailto:timo.haeckel@haw-hamburg.de}{timo.haeckel}, \href{mailto:philipp.meyer@haw-hamburg.de}{philipp.meyer}, \href{mailto:mehmet.mueller@haw-hamburg.de}{mehmet.mueller}, \href{mailto:jan.schmitt-solbrig@haw-hamburg.de}{jan.schmitt-solbrig}, \href{mailto:franz.korf@haw-hamburg.de}{franz.korf}, \href{mailto:t.schmidt@haw-hamburg.de}{t.schmidt}\}@haw-hamburg.de}
}

\maketitle

\setlength{\TPHorizModule}{\paperwidth}
\setlength{\TPVertModule}{\paperheight}
\TPMargin{5pt}
\begin{textblock}{0.8}(0.1,0.02)
     \noindent
     \footnotesize
     If you cite this paper, please use the original reference:
     Timo H\"ackel, Philipp Meyer, Mehmet Mueller, Jan Schmitt-Solbrig, Franz Korf, Thomas C. Schmidt. 
     ``Dynamic Service-Orientation for Software-Defined In-Vehicle Networks,'' In: \emph{Proceedings of the IEEE 97th Vehicular Technology Conference (VTC2023-Spring)}. IEEE, June 2023.
\end{textblock}

\begin{abstract}
    Modern In-Vehicle Networks (IVNs) are composed of a large number of devices and services linked via an Ethernet-based time-sensitive network.
    Communication in future IVNs will become more dynamic as services can be updated, added, or removed during runtime.
    This requires a flexible and adaptable IVN, for which Software-Defined Networking (SDN) is a promising candidate. 
    In this paper, we show how SDN can be used to support a dynamic, service-oriented network architecture. 
    We demonstrate our concept using the SOME/IP protocol, which is the most widely deployed implementation of automotive service-oriented architectures. 
    In a simulation study, we evaluate the performance of SOME/IP-adaptive SDN control compared to standard Ethernet switching and non-optimized SDN.
    Our results show an expected overhead introduced by the central SDN controller, which is, however, reduced by up to 50\% compared to SOME/IP-unaware SDN.
    For a large number of services, the setup time is in the order of milliseconds, which matches standard Ethernet switching.
    A SOME/IP-aware SDN controller can optimize the service discovery to improve adaptability, robustness, security, and Quality-of-Service of the IVN while remaining transparent to existing SOME/IP implementations.
\end{abstract}

\begin{IEEEkeywords}
    SDN, In-Vehicle Networks, Service-Oriented Architectures, SOME/IP, Service Discovery
\end{IEEEkeywords}

\begin{acronym}
	\acro{ACC}[ACC]{Adaptive Cruise Control}
	\acro{ACDC}[ACDC]{Automotive Cyber Defense Center}
	\acro{ACL}[ACL]{Access Control List}
	\acro{ADS}[ADS]{Anomaly Detection System}
	\acroplural{ADS}[ADSs]{Anomaly Detection Systems}
	\acro{ADAS}[ADAS]{Advanced Driver Assistance Systems}
	\acro{API}[API]{Application Programming Interface}
	\acro{AVB}[AVB]{Audio Video Bridging}
	\acro{ARP}[ARP]{Address Resolution Protocol}
	\acro{BE}[BE]{Best-Effort}
	\acro{CAN}[CAN]{Controller Area Network}
	\acro{CBM}[CBM]{Credit Based Metering}
	\acro{CBS}[CBS]{Credit Based Shaping}
	\acro{CNC}[CNC]{Central Network Controller}
	\acro{CMI}[CMI]{Class Measurement Interval}
	\acro{CoRE}[CoRE]{Communication over Realtime Ethernet}
	\acro{CT}[CT]{Cross Traffic}
	\acro{CM}[CM]{Communication Matrix}
	\acro{DoS}[DoS]{Denial of Service}
	\acro{DDoS}[DDoS]{Distributed Denial of Service}
	\acro{DDS}[DDS]{Data Distribution Service}
	\acro{DPI}[DPI]{Deep Packet Inspection}
	\acro{ECU}[ECU]{Electronic Control Unit}
	\acroplural{ECU}[ECUs]{Electronic Control Units}
	\acro{FDTI}[FDTI]{Fault Detection Time Interval}
	\acro{FHTI}[FHTI]{Fault Handling Time Interval}
	\acro{FRTI}[FRTI]{Fault Reaction Time Interval}
	\acro{FTTI}[FTTI]{Fault Tolerant Time Interval}
	\acro{GCL}[GCL]{Gate Control List}
	\acro{HTTP}[HTTP]{Hypertext Transfer Protocol}
	\acro{HMI}[HMI]{Human-Machine Interface}
	\acro{HPC}[HPC]{High-Performance Controller}
	\acro{IA}[IA]{Industrial Automation}
	\acro{IDS}[IDS]{Intrusion Detection System}
	\acroplural{IDS}[IDSs]{Intrusion Detection Systems}
	\acro{IEEE}[IEEE]{Institute of Electrical and Electronics Engineers}
	\acro{IGMP}[IGMP]{Internet Group Management Protocol}
	\acro{IoT}[IoT]{Internet of Things}
	\acro{IP}[IP]{Internet Protocol}
	\acro{ICT}[ICT]{Information and Communication Technology}
	\acro{IVNg}[IVN]{In-Vehicle Networking}
	\acro{IVN}[IVN]{In-Vehicle Network}
	\acroplural{IVN}[IVNs]{In-Vehicle Networks}
	\acro{LIN}[LIN]{Local Interconnect Network}
	\acro{MOST}[MOST]{Media Oriented System Transport}
	\acro{NADS}[NADS]{Network Anomaly Detection System}
	\acroplural{NADS}[NADSs]{Network Anomaly Detection Systems}
	\acro{OEM}[OEM]{Original Equipment Manufacturer}
	\acro{OTA}[OTA]{Over-the-Air}
	\acro{P4}[P4]{Programming Protocol-independent Packet Processors}
	\acro{PCP}[PCP]{Priority Code Point}
	\acro{RC}[RC]{Rate-Constrained}
	\acro{REST}[ReST]{Representational State Transfer}
	\acro{RPC}[RPC]{Remote Procedure Call}
	\acro{SD}[SD]{Service Discovery}
	\acro{SDN}[SDN]{Software-Defined Networking}
	\acro{SDN4CoRE}[SDN4CoRE]{Software-Defined Networking for Communication over Real-Time Ethernet}
	\acro{SIEM}[SIEM]{Security Information and Event Management}
	\acro{SOA}[SOA]{Service-Oriented Architecture}
	\acro{SOC}[SOC]{Security Operation Center}
	\acro{SOME/IP}[SOME/IP]{\textit{Scalable service-Oriented MiddlewarE over IP}}
	\acro{SR}[SR]{Stream Reservation}
	\acro{SRP}[SRP]{Stream Reservation Protocol}
	\acro{SW}[SW]{Switch}
	\acroplural{SW}[SW]{Switches}
	\acro{TAS}[TAS]{Time-Aware Shaping}
	\acro{TCP}[TCP]{Transmission Control Protocol}
	\acro{TDMA}[TDMA]{Time Division Multiple Access}
	\acro{TSN}[TSN]{Time-Sensitive Networking}
	\acro{TSSDN}[TSSDN]{Time-Sensitive Software-Defined Networking}
	\acro{TT}[TT]{Time-Triggered}
	\acro{TTE}[TTE]{Time-Triggered Ethernet}
	\acro{TTL}[TTL]{Time to Live}
	\acro{UDP}[UDP]{User Datagram Protocol}
	\acro{UN}[UN]{United Nations}
	\acro{QoS}[QoS]{Quality-of-Service}
	\acro{V2X}[V2X]{Vehicle-to-X}
	\acro{WS}[WS]{Web Services}
	\acro{ZC}[ZC]{Zone Controller}

\end{acronym}


\section{Introduction}%
\label{sec:introduction}
Today's vehicles accommodate a large number of interconnected \acp{ECU}, the software of which is expected to undergo 
faster development cycles with frequent updates and service reconfigurations. 
The introduction of a \ac{SOA} promises to support this increasing dynamic~\cite{kobct-osoas-17}.
The \ac{IVN} plays a fundamental role in the performance, safety, and security of these interconnected services~\cite{mrfm-svcjr-21}. 
Ethernet emerges as the next generation \ac{IVN} technology to extend capacity and flexibility, while replacing existing bus systems.
\ac{TSN} enhances Ethernet with \ac{QoS} features, e.g., deterministic communication and redundancy.

One major challenge of future \acp{IVN} is to transform from statically pre-configured into service-oriented networks that support dynamic service adaptation.
With its centralized control plane, \ac{SDN} promises a dynamic and flexible \ac{IVN}~\cite{hhlng-saeea-20}. 
In previous work, we presented \ac{TSSDN} that supports \ac{QoS} and security requirements of in-vehicle applications~\cite{hmks-snsti-19,hmks-stsnv-23}.
An SDN controller can reconfigure the network according to service availability, including updates or failures.
\ac{TSN} resource partitioning~\cite{lbp-bptjr-21} allows adding new traffic flows dynamically without affecting a static configuration defined at design time.
Additionally, failover mechanisms and seamless service mobility~\cite{erf-sbrjr-21} can improve the robustness of the \ac{IVN}.

Evolving automotive systems is challenging, as industry practices and backward interoperability need to be met at reasonable overhead~\cite{mrfm-svcjr-21}.
The \ac{SOME/IP} is a widely used protocol for automotive \acp{SOA} standardized by AUTOSAR~\cite{autosar-someip-22}.
\ac{SOME/IP} offers \ac{SD}~\cite{autosar-someip-sd-22} as a complementary service.
SDN-supported SOA in vehicles opens powerful potentials.
Options range from discovery optimizations to \ac{QoS}, security, and robustness improvements.
To support in-car \ac{SOA}, the \ac{SDN} controller must know all services on the \ac{IVN}, and thus support automotive protocols for service discovery.

In this work, we present a network control scheme for the \ac{SOME/IP} \ac{SD} based on the \ac{SDN} paradigm. 
We design an \ac{SDN} controller application that fully supports \ac{SOME/IP} \ac{SD}.
It intercepts discovery messages, learns about services, directly responds to requests, and sets up paths automatically.
Our approach is completely transparent to existing \ac{SOME/IP} implementations.
We discuss further potentials of supporting \ac{SOME/IP} communication with SDN (service discovery optimization, service mobility, \ac{QoS} improvements, discovery protection).
In simulation, we compare the scalability of our approach to non-optimized SDN and standard Ethernet.

The remainder of this work is structured as follows:
In Sec.~\ref{sec:background_and_related_work}, we discuss related work.
Sec.~\ref{sec:concept} introduces the \ac{SOME/IP} \ac{SD} mechanism, the \ac{SDN} paradigm, our methodology to support automotive \acp{SOA} with \ac{SDN}, and further potentials enabled by their combination. 
Sec.~\ref{sec:eval} evaluates the performance of the proposed \ac{SDN} supported \ac{SOME/IP} \ac{SD}.
Finally, Sec.~\ref{sec:conclusion_and_outlook} concludes this work and outlines future work.


\section{Background and Related Work}
\label{sec:background_and_related_work}
Today, an \ac{IVN} is a complex distributed system with a multitude of devices and services communicating via a combination of bus systems and switched Ethernet.
Combined with \acf{TSN}, Automotive Ethernet is a promising candidate for the backbone of next generation \acp{IVN}.
Gateways that translate between different networks (e.g., CAN and Ethernet) and protocols are commonly used to enable interoperability and backward compatibility~\cite{iks-aiejr-22}.

\subsection{Service-Oriented Architectures in Vehicles}
\ac{SOA} is proposed to enable flexible and dynamic application placement, and frequent updates in automotive networks~\cite{kobct-osoas-17}.
In-vehicle applications can act as service providers to make certain functions and data available to consuming applications.
While dynamic service discovery can improve flexibility of, for example, navigation, infotainment, diagnostics, and driver assistance services, safety-critical services may still require dedicated, static connections to ensure reliability~\cite{chrmk-qaasc-19}.

Two major candidates are considered as the communication protocol for \acp{SOA} in vehicles~\cite{iks-aiejr-22}:
(1) \ac{SOME/IP}~\cite{autosar-someip-22} is explicitly tailored to the automotive environment and enables service-oriented communication via TCP/UDP-IP.
(2) \ac{DDS}~\cite{omg-dds-15} from the Object Management Group is a viable alternative available in the AUTOSAR platform, but not designed for automotive applications and not widely used by automotive companies~\cite{iks-aiejr-22}.
This work focuses on \ac{SOME/IP} due to its widespread deployment in the automotive domain.
Nevertheless, our work translates to other protocols such as \ac{DDS}.

In previous work~\cite{chrmk-qaasc-19}, we assessed the design space of vehicular services and proposed a mechanism to enable \ac{QoS} within a vehicular middleware.
Since rollout of \ac{SOME/IP}-\ac{SD} in AUTOSAR 4.1, \ac{SOA} is a standard feature for future \acp{IVN}.
Kampmann et al.~\cite{kakww-dsosa-19} propose containerized services to be placed and activated on dynamic allocated hardware resources during runtime.
The dynamic nature of \acp{SOA} poses challenges to a traditionally pre-configured \ac{IVN}, as it must adapt to changes in service availability during runtime.

\subsection{Supporting Automotive Network Functions with SDN}
\acf{SDN} has the potential to increase the flexibility and performance of networks~\cite{mabpp-oeicn-08}, in particular in well-known environments. 
\ac{SDN} centralizes the control plane and separates it from the data plane, allowing a central controller to perform network functions such as routing, firewalling and load balancing for the local network.

In cars, \ac{SDN} can enable a reconfigurable and flexible network architecture that adapts to changes in the network, e.g., software updates and downloadable drive assistance systems~\cite{hhlng-saeea-20}.
In previous work~\cite{hmks-snsti-19,hmks-stsnv-23}, we presented \ac{TSSDN}, an integration of SDN with TSN, and showed how \ac{SDN} can significantly enhance \ac{IVN} security.
Ergen\c{c} et al.~\cite{erf-sbrjr-21}, illustrate service-based resilience for \acp{IVN} by configuring backup nodes for critical services.
In case of failures, those can be activated, which also requires changes in the network configuration. 

Intercepting packets of network control protocols is a common approach to optimize networking objectives via \ac{SDN}.
Examples include the management of the \ac{ARP}~\cite{ap-saejr-16} or IP multicast routing~\cite{ima-smsjr-18}.
Bertaux et al.~\cite{bhmba-dbcjr-14} present a first design for an \ac{SDN} application that dynamically allocates network resources for \ac{DDS} applications.
Such a mechanism is missing in \ac{SOME/IP} and could enable the \ac{IVN} to adapt to changes during runtime.

In this work, we present a concept for an \ac{SDN} controller application that fully supports the \ac{SOME/IP} \ac{SD} protocol. 
It can intercept and handle service announcements and subscriptions to manage network resources but remains fully transparent for existing \ac{SOME/IP} implementations and applications.


\section{SDN-Supported SOME/IP Service Discovery}
\label{sec:concept}
Our concept applies the SDN paradigm to service-oriented in-vehicle communication using the \ac{SOME/IP} protocol.
First, we introduce the \ac{SOME/IP} \ac{SD} mechanism, next our methodology to support automotive \acp{SOA} with \ac{SDN}, and finally cover \ac{SOME/IP} message handling in a controller application.

\subsection{The \acs{SOME/IP} Service Discovery}
\label{subsec:someip}
\begin{figure}
    \centering
    \includegraphics[width=1.0\linewidth]{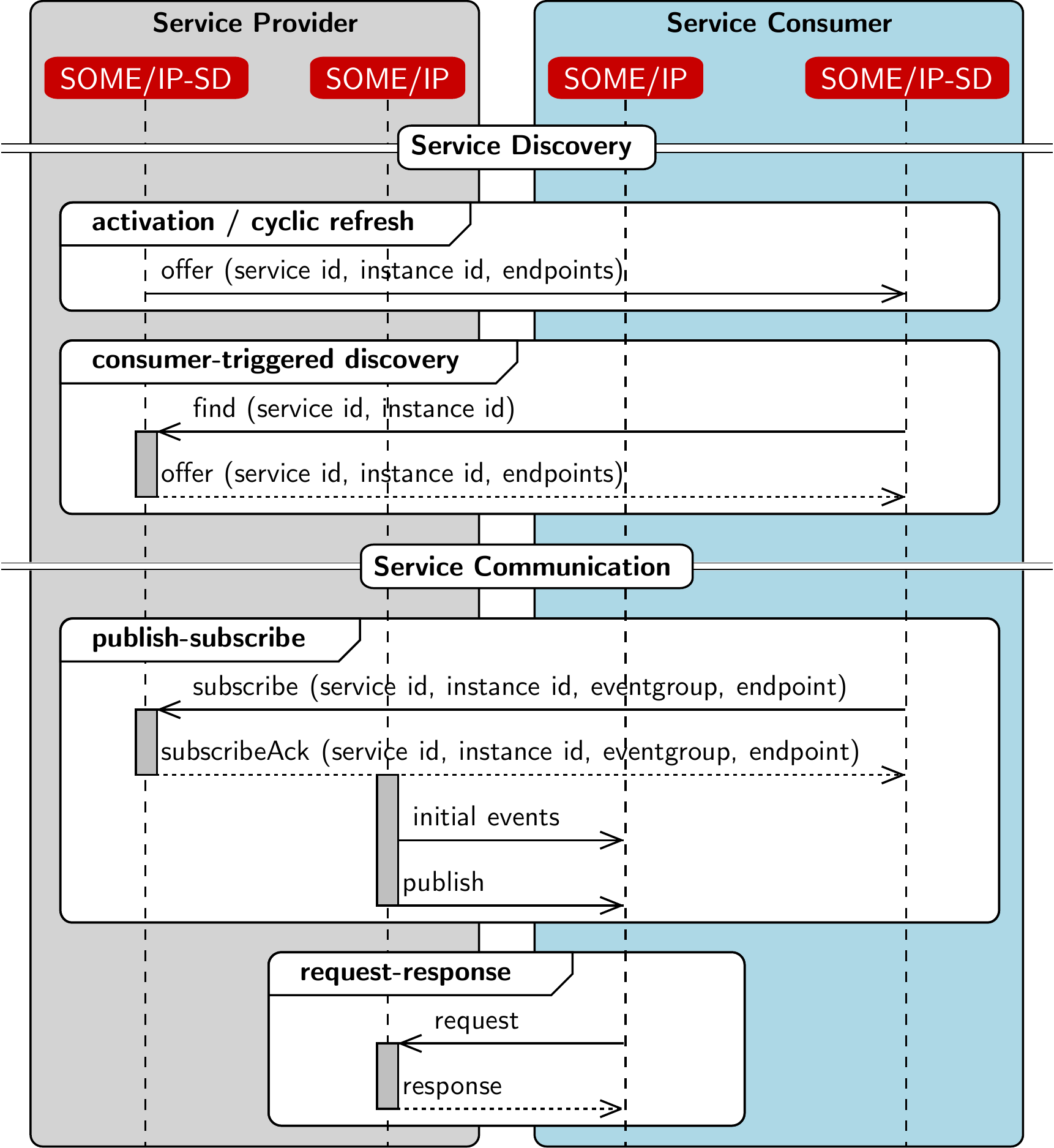}
    \caption{\acs{SOME/IP} service discovery and communication.}
    \label{fig:someip_uml}
    \vspace{-2pt}
\end{figure}

The AUTOSAR \ac{SOME/IP} protocol~\cite{autosar-someip-22} includes a \acf{SD} protocol~\cite{autosar-someip-sd-22} with an offline-defined endpoint (IP multicast group / UDP port) on each device.
Fig.~\ref{fig:someip_uml} shows a sequence with \ac{SOME/IP} \ac{SD} and service communication.

\ac{SOME/IP} has two message types for service discovery: \codeword{find} to request a service, and \codeword{offer} to announce a service.
Providers send an \codeword{offer} to the \ac{SD} multicast group upon service activation or as a cyclic refresh.
A consumer can initiate a \codeword{find}-request for a service as a \ac{SD} multicast.
Multiple instances of the same service can coexist in a vehicle.
All providers of the requested service respond with an unicast \codeword{offer}.
A \ac{SOME/IP} service identifies with service ID, instance ID, and major/minor version.
The instance ID, major/minor version can be wildcarded in the \codeword{find}-request if any instance of the specified service ID is applicable.

Discovered services can communicate following a publish-subscribe or a request-response model.
For the latter, a consumer requests a data field or method from a provider via unicast each time the data is needed and awaits the response.
In the publish-subscribe model, a consumer subscribes once to a service instance via unicast to be notified repeatedly. 
The provider acknowledges the subscription and communication endpoints are created. 
The provider can then send initial events and thereafter publish updates on-change or periodically.

A \ac{TTL} field is used to limit the lifetime of an \codeword{offer}, \codeword{find}, or \codeword{subscribe}.
To withdraw an offered service or subscription, the message is sent with identical service ID and instance ID, but with the \ac{TTL} set to 0.

\subsection{Architecture of an SDN-Supported Automotive SOA} 
\begin{figure}
    \centering
    \includegraphics[width=0.9\linewidth,trim={0.7cm 0.8cm 0.9cm 0.8cm},clip=true]{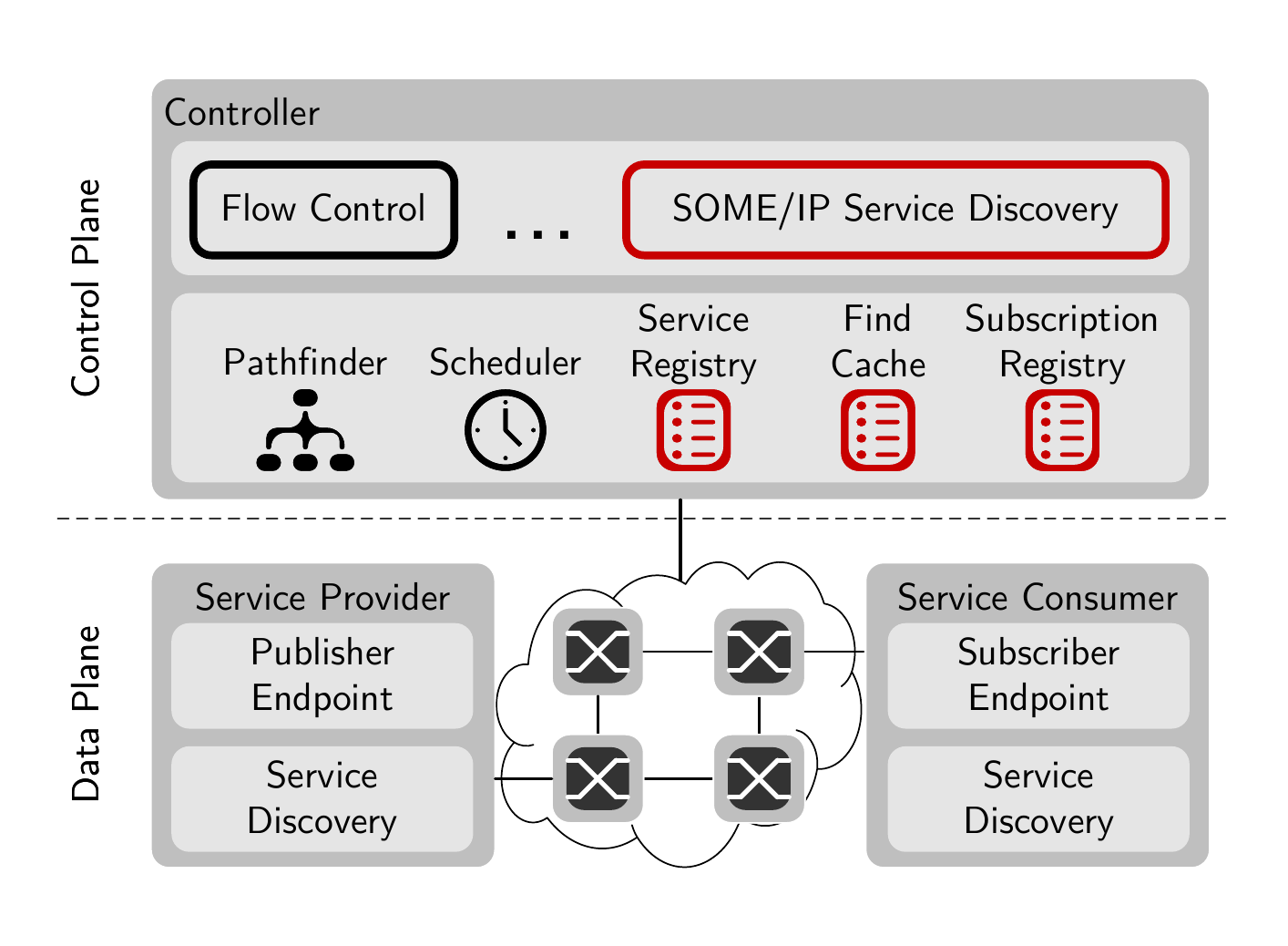}
    \caption{Concept for an SDN-supported automotive SOA. Components in red are added, others remain unaltered.}
    \label{fig:sdn_concept}
\end{figure}

We integrate the \ac{SOME/IP} service management with \ac{SDN}, enabling the SDN controller to gather information about \ac{SOME/IP} services and exploit it for network optimization. 
Fig.~\ref{fig:sdn_concept} gives an architectural overview of the proposed concept.

The network is divided into control plane and data plane following the \ac{SDN} paradigm.
On the data plane, switches connect service providers and consumers and forward packets between endpoints based on their forwarding tables.
The \ac{SDN} controller on the control plane manages the network and the forwarding tables of network devices.

We add a specialized network application to the SDN controller that intercepts all \ac{SOME/IP} \ac{SD} messages and extracts information about the services. 
It stores  this information in three \ac{SD} tables:
(1) A service registry that contains all known service instances and their endpoints.
(2) A find cache stores all open \codeword{find}-requests which the controller can only answer after the service was discovered.
(3) A subscription registry tracks subscriptions and their status (e.g., active or subscription requested) for the publish-subscribe model.

\subsection{Controlling the SOME/IP Service Discovery with SDN}

\begin{table}
    \centering
    \caption{
        Handling SOME/IP service discovery messages on the control plane depending on the state -- \textit{known} when in the service registry, \textit{requested} when in the find cache, and \textit{subscribed} when in the subscription registry.
    }
    \label{tab:message_handling}
    \setlength{\tabcolsep}{3pt}
    \begin{tabularx}{\linewidth}{l l p{0.46\linewidth}}
        \toprule
        Message & State \texttt{\textcolor{CoreBlack}{(X)}} & Reaction \\
        \midrule 
        \multirow{2}{*}{\texttt{\textcolor{CoreBlack}{find(X)}}} & \textit{Known} & respond with \texttt{\textcolor{CoreBlack}{offer}} \\
         & Not \textit{known} & Send \texttt{\textcolor{CoreBlack}{find}} to multicast group \\
        \cmidrule(rl){1-3}
        \multirow{4}{*}{\texttt{\textcolor{CoreBlack}{offer(X)}}} & \textit{Requested} & Update service registry, forward to requester or multicast group \\
         & Not \textit{requested} & Update service registry, forward to multicast group  \\
        \cmidrule(rl){1-3}
        \multirow{3}{*}{\texttt{\textcolor{CoreBlack}{subscribe(X)}}} & \textit{Known} & Update subscription registry, forward to provider \\
         & Not \textit{known} & Send negative acknowledgement \\ 
        \cmidrule(rl){1-3}
        \multirow{3}{*}{\texttt{\textcolor{CoreBlack}{subscribeAck(X)}}} & \textit{Subscribed} & Update subscription registry, install forwarding,  forward to subscriber \\
         & Not \textit{subscribed} & No handling \\
        \bottomrule
    \end{tabularx}
\end{table}

Table~\ref{tab:message_handling} shows how the controller application handles \ac{SOME/IP} \ac{SD} messages depending on the state of the three \ac{SD} tables. 
The controller responds directly to \codeword{find} messages for registered services and sends all known instances of a service ID when the instance ID is a wildcard.
If the service instance is not yet known, the \codeword{find} is cached and forwarded to the multicast group.
On \codeword{offer} messages, the controller updates the service registry.
If the \codeword{offer} is sent as a direct reply to a cached \codeword{find} it is forwarded to the requester, otherwise to the multicast group.

All \codeword{subscribe} messages are forwarded to the provider.
If the service is known, they are added to the subscription registry.
When the controller receives a \codeword{subscribeAck} and the subscription is in the table, it installs a forwarding rule.
Then it forwards the message to the subscriber.
For subscriptions that are already active (e.g., in case of multicast), the controller updates existing rules along the path.

Withdrawn offers are updated in the subscription and service registry and forwarded to the multicast group.
Withdrawn subscriptions are updated in the subscription registry and forwarded to the service provider.
Previously installed rules must also be removed or updated in forwarding devices.

These mechanisms are fully transparent towards the endpoints and do not require any changes in the \ac{SOME/IP} \ac{SD} protocol implementation nor the applications 
Introducing a central network controller, however, impacts service discovery performance, since all \ac{SD} messages are forwarded to the controller, which we evaluate in Section~\ref{sec:eval}.

\subsection{Potentials of SDN-Supported Automotive SOA}

Further potentials emerge if the \ac{SDN} controller is aware of the \ac{SOME/IP} \ac{SD} protocol:
\subsubsection{Optimized service discovery}
The controller can automatically install flows and efficiently add new subscribers to multicast flows after receiving a \codeword{subscribeAck} message.
Furthermore, load on the data plane could be reduced by directly forwarding \ac{SD} messages to the known \ac{SOME/IP} endpoints, skipping links between forwarding devices.
With \ac{SDN}, any Ethernet topologies such as rings can be supported for \ac{SOME/IP} services, enabling load balancing and redundancy.

\subsubsection{Seamless service mobility}
The controller can seamlessly reconfigure publishers and subscribers, as they move from one device to another without disrupting communications.
In addition, it can hand over subscriptions from one publisher instance to another instance of the same service if the original publisher fails.
Other reconfiguration mechanisms could improve the robustness of the \ac{IVN}, such as the fast handover in case of a link failure.

\subsubsection{\acl{QoS} enforcement}
The \ac{IVN} has strict QoS requirements with particular real-time capabilities.
In traditional networks, it can be challenging to translate QoS requirements of application layer services onto the underlying link layer.
Already on host devices, preconfigured mapping from network layer QoS options to link layer QoS options is required.
Communicating such requirements to the network can be even more challenging. 
In the proposed approach, the controller can enforce QoS requirements of the subscriber in the network.
In previous work, we have shown how a central controller can configure the \ac{TSN}-scheduling of switches~\cite{hmks-stsnv-23}.

\subsubsection{Discovery protection}
While IVN security is imperative for ensuring vehicle safety, in-vehicle communication protocols and architectures often lack security mechanisms~\cite{mrfm-svcjr-21}.
In the case of \ac{SOME/IP}, end-to-end encryption can be used to protect the message payload. 
However, the discovery of \ac{SOME/IP} services is not protected, and malicious nodes could easily spoof the discovery messages, e.g., to announce conflicting publisher service instances or add additional subscriptions to increase the network load.
The controller can support security mechanisms for \ac{SOME/IP} \ac{SD} to verify the authenticity of discovery messages and control access policies for providers and consumers.


\section{Performance analysis}
\label{sec:eval}
Our study compares the performance of the presented concept compared to non-optimized SDN forwarding and standard Ethernet switches in simulation.
Our key performance metric is the time to set up all subscriptions in a network, from first producer until the last subscription is established.

\subsection{Simulation Environment}

Our simulation environment is based on the OMNeT++ simulator~\cite{omnet-inet} with the INET framework, and the OpenFlowOMNeTSuite~\cite{kj-oeojr-13}.
For our implementation, we use the CoRE4INET, SDN4CoRE~\cite{hmks-smsdn-19}, and SOA4CoRE~\cite{chrmk-qaasc-19} frameworks which our research group maintains.
They are open source on \mbox{\textit{\url{github.com/CoRE-RG}}}.
SOA4CoRE implements the \ac{SOME/IP} and \ac{SOME/IP} \ac{SD} protocol, and our proposed controller application for \ac{SOME/IP} \ac{SD} is added to SDN4CoRE.

\subsection{Evaluation Scenario}
\begin{figure}
    \centering
    \includegraphics[width=1.0\linewidth, trim=0.62cm 0.62cm 0.62cm 0.72cm, clip=true]{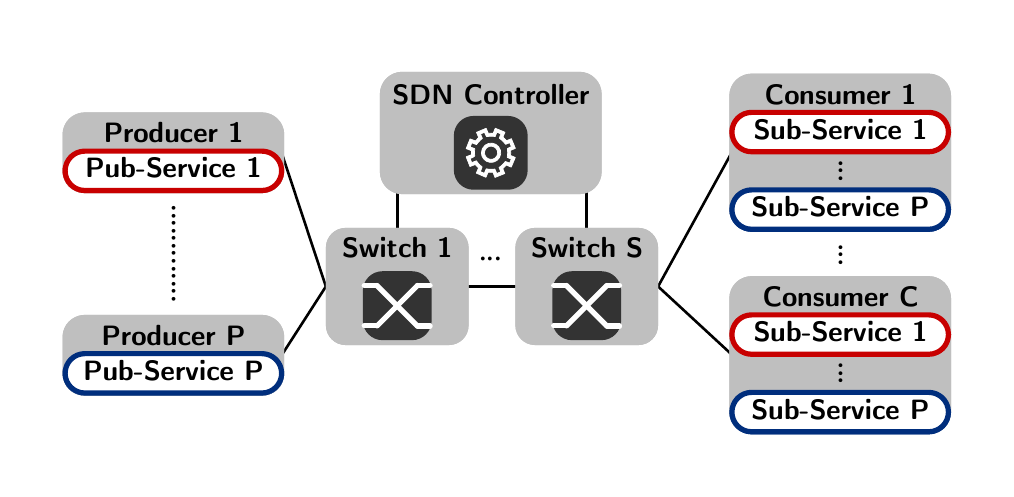}
    \caption{Topology for the scalability comparison with 1 to 5 switches (S), 1 to 50 producers (P), and 1 to 50 consumer nodes (C) with one subscriber service per producer (P).}
    \label{fig:scalingtopo}
    \vspace{-7pt}
\end{figure}

Fig.~\ref{fig:scalingtopo} shows the topology used for the comparison.
It consists of switches, producer nodes, and consumer nodes connected via \SI{1}{\giga\bit\per\second} Ethernet links.
All switches are connected to a controller in the SDN variants.

To investigate scalability, we vary the number of producer (P) and consumer nodes (C) from 1 to 50, and the number of switches between them from 1 to 5, since we expect that this number is not exceeded in a real \ac{IVN}.
Each producer has one publisher service.
Each consumer has one subscriber service per publisher.
We simulate all 192 parameter combinations and measure the time to set up all subscriptions.

The simulation models provide a wide range of configuration parameters.
Switches have a hardware forwarding delay of \SI{8}{\micro\second}.
The SDN controller uses the OpenFlow protocol to configure the switches.
The OpenFlow messages processing time in the controller application is \SI{100}{\micro\second}, based on the worst-case performance of the best-performing controller implementation we have evaluated in previous work~\cite{rhmks-rapesc-20}.
We assume the switch processing time is similar and set it to \SI{100}{\micro\second}, but could not find any data in the literature on the performance of OpenFlow processing on switches.
The controller and the switches can handle multiple OpenFlow packets in parallel.

\subsection{Results}

\begin{figure*}
	\centering
    \begin{tikzpicture}
        \begin{axis}[width=.32\linewidth, height=.28\linewidth,
            title={1 switch},
            every axis title/.style={below right,at={(0,1)},draw=black,fill=white},
            change y base,
            y unit=\second,
			ylabel=log(startup time),
			xlabel={Producers},
            x unit=\#,
            xmin=1, xmax=50,
            ymin=0.00001, ymax=0.1,
            ymode=log,
            xtick={1,10,20,30,40,50},
            ytick={0.00001,0.0001,0.001,0.01,0.1,1,10},
            xtick pos=bottom,
            ]
            \addplot [CoreGreen, dashed, thick] table [x, y, col sep=comma] {./data/scaling/ethernet_S=1_C=1.csv};\label{plots:eth_1}
            \addplot [CoreRed, dashed, thick] table [x, y, col sep=comma] {./data/scaling/parallel_100c_100s_sdn_S=1_C=1.csv};\label{plots:sdn_1}
            \addplot [CoreBlue, dashed, thick] table [x, y, col sep=comma] {./data/scaling/parallel_100c_100s_vanilla_S=1_C=1.csv};\label{plots:vanilla_1}
            \addplot [CoreGreen, thick] table [x, y, col sep=comma] {./data/scaling/ethernet_S=1_C=50.csv};\label{plots:eth_50}
            \addplot [CoreRed, thick] table [x, y, col sep=comma] {./data/scaling/parallel_100c_100s_sdn_S=1_C=50.csv};\label{plots:sdn_50}
            \addplot [CoreBlue, thick] table [x, y, col sep=comma] {./data/scaling/parallel_100c_100s_vanilla_S=1_C=50.csv};\label{plots:vanilla_50}
        \end{axis}
    \end{tikzpicture}
    \begin{tikzpicture}
        \begin{axis}[width=.32\linewidth, height=.28\linewidth,
            title={2 switches},
            every axis title/.style={below right,at={(0,1)},draw=black,fill=white},
            change y base,
			xlabel={Producers},
            x unit=\#,
            xmin=1, xmax=50,
            ymin=0.00001, ymax=0.1,
            ymode=log,
            xtick={1,10,20,30,40,50},
            ytick={0.00001,0.0001,0.001,0.01,0.1,1,10},
            xtick pos=bottom,
            ]
            \addplot [CoreGreen, dashed, thick] table [x, y, col sep=comma] {./data/scaling/ethernet_S=2_C=1.csv};
            \addplot [CoreRed, dashed, thick] table [x, y, col sep=comma] {./data/scaling/parallel_100c_100s_sdn_S=2_C=1.csv};
            \addplot [CoreBlue, dashed, thick] table [x, y, col sep=comma] {./data/scaling/parallel_100c_100s_vanilla_S=2_C=1.csv};
            \addplot [CoreGreen, thick] table [x, y, col sep=comma] {./data/scaling/ethernet_S=2_C=50.csv};
            \addplot [CoreRed, thick] table [x, y, col sep=comma] {./data/scaling/parallel_100c_100s_sdn_S=2_C=50.csv};\label{plots:sdn_50}
            \addplot [CoreBlue, thick] table [x, y, col sep=comma] {./data/scaling/parallel_100c_100s_vanilla_S=2_C=50.csv};
        \end{axis}
    \end{tikzpicture}
    \begin{tikzpicture}
        \begin{axis}[width=.32\linewidth, height=.28\linewidth,
            title={5 switches},
            every axis title/.style={below right,at={(0,1)},draw=black,fill=white},
            change y base,
			xlabel={Producers},
            x unit=\#,
            xmin=1, xmax=50,
            ymin=0.00001, ymax=0.1,
            ymode=log,
            xtick={1,10,20,30,40,50},
            ytick={0.00001,0.0001,0.001,0.01,0.1,1,10},
            xtick pos=bottom,
            ]
            \addplot [CoreGreen, dashed, thick] table [x, y, col sep=comma] {./data/scaling/ethernet_S=5_C=1.csv};
            \addplot [CoreRed, dashed, thick] table [x, y, col sep=comma] {./data/scaling/parallel_100c_100s_sdn_S=5_C=1.csv};
            \addplot [CoreBlue, dashed, thick] table [x, y, col sep=comma] {./data/scaling/parallel_100c_100s_vanilla_S=5_C=1.csv};
            \addplot [CoreGreen, thick] table [x, y, col sep=comma] {./data/scaling/ethernet_S=5_C=50.csv};
            \addplot [CoreRed, thick] table [x, y, col sep=comma] {./data/scaling/parallel_100c_100s_sdn_S=5_C=50.csv};\label{plots:sdn_50}
            \addplot [CoreBlue, thick] table [x, y, col sep=comma] {./data/scaling/parallel_100c_100s_vanilla_S=5_C=50.csv};
        \end{axis}
    \end{tikzpicture}
    \begin{tikzpicture}
        \hspace{8pt}
        \matrix[
            font=\footnotesize,
            matrix of nodes,
            align=left,
            anchor=west,
            inner sep=0.1em,
            draw
            ]
            {
                \ref{plots:eth_1}&{\hspace{-1pt}w/o SDN, \hspace{4pt}1 consumer \hspace{3pt}per producer}&[5pt]
                \ref{plots:sdn_1}&{\hspace{-1pt}SDN optimized, \hspace{4pt}1 consumer \hspace{3pt}per producer}&[5pt]
                \ref{plots:vanilla_1}&{\hspace{-1pt}SDN vanilla, \hspace{4pt}1 consumer \hspace{3pt}per producer}\\
                \ref{plots:eth_50}&{w/o SDN, 50 consumers per producer}&[5pt]
                \ref{plots:sdn_50}&{SDN optimized, 50 consumers per producer}&[5pt]
                \ref{plots:vanilla_50}&{SDN vanilla, 50 consumers per producer}\\
            };
    \end{tikzpicture}
    \caption{
        Comparison of the presented approach (SDN optimized) to non-optimized SDN forwarding (SDN vanilla) and standard Ethernet switches (w/o SDN) in terms of scalability.
        The logarithmic scale depicts the total time to set up all subscriptions for an increasing number of producers with 1 and 50 consumers per producer.
    }
    \label{fig:ethvsSDNovsSDNv}
    \vspace{-8pt}
\end{figure*}
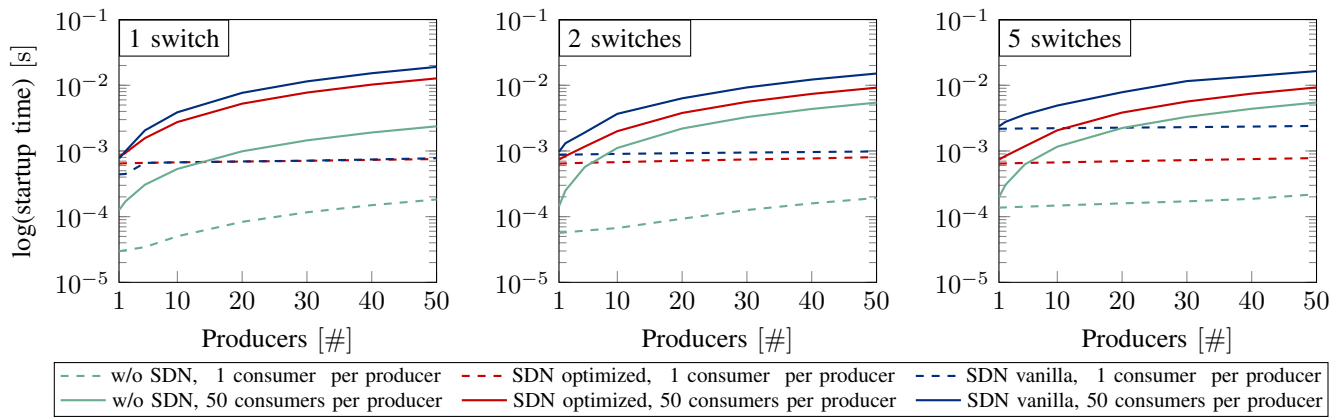 

Fig.~\ref{fig:ethvsSDNovsSDNv} compares the setup time of the presented approach (SDN optimized), non-optimized SDN forwarding (SDN vanilla), and standard Ethernet (w/o SDN) for 1 to 50 producers on 1, 2, and 5 switches on a logarithmic scale. 
For simplicity, the graphs only depict results for 1 and 50 consumers per producer, the others show a similar trend.

In all cases, we observe an approximately linear increase in setup time with the number of producers for all three approaches.
Compared to standard Ethernet switching, the SDN solutions are about one order of magnitude slower, which is to be expected due to the delay caused by forwarding each \ac{SOME/IP} \ac{SD} packet to the central \ac{SDN} controller. 
This delay is highly dependent on the OpenFlow processing time.
Optimized controllers, and switches might come closer to the Ethernet performance.
Nevertheless, the additional delay only affects the setup time of the subscriptions and not the actual data transfer of the service.

For setting up one subscription over 5 switches, the optimized SDN approach takes \SI{0.6}{\milli\second}, non-optimized SDN \SI{2.2}{\milli\second} and the non-SDN variant only \SI{0.1}{\milli\second}.
This is a good indicator that the time to migrate a service from one node to another is less than \SI{1}{\milli\second}.
The setup of subscriptions for 50 publishers with 50 subscribers each (2500 subscriptions), is considered a cold start as all services announce their availability and their required subscriptions at the same time.
The setup for all connections takes about \SI{5.5}{\milli\second} without SDN, \SI{16.4}{\milli\second} with vanilla SDN, and \SI{9.3}{\milli\second} with our optimized SDN approach. 

Overall, the presented approach of a SOME/IP-aware SDN controller improves performance by up to 50\% compared to a non-optimized SDN.
Although a migration time of \SI{1}{\milli\second} is acceptable for most in-vehicle services, it may not be acceptable for safety-critical services, e.g., collision detection, which likely require configured routes and redundancy.
Most services, however, are discovered when the vehicle is started.
The fastest required service availability in current cars is about \SI{200}{\milli\second} and even lower for, e.g., infotainment services.
Therefore, the setup time for all in-vehicle connections (2500 services in less than \SI{10}{\milli\second}) is acceptable for all kinds of services, including safety-critical services. 


\section{Conclusion and Outlook}%
\label{sec:conclusion_and_outlook}

This paper proposed an SDN-based network control scheme for \ac{SOME/IP} \ac{SD}.
We designed an \ac{SDN} controller application that fully supports \ac{SOME/IP} \ac{SD}, while remaining transparent to existing \ac{SOME/IP} implementations.
The controller detects available service instances and automatically sets up paths in forwarding devices for acknowledged subscriptions.

We evaluated the performance of our approach in a simulation-based study, comparing its scalability to non-optimized SDN and standard Ethernet switching. 
Our approach improved the SDN performance by up to 50\% compared to the non-optimized SDN solution.
The central controller processing all \ac{SOME/IP} \ac{SD} messages significantly reduces SDN performance, but the additional delay only affects subscription setup time, not the actual data transfer.
Our approach achieves a setup time below \SI{10}{\milli\second} for 2500 subscriptions, complying with the fastest service availability normally required in a vehicle, which is about \SI{200}{\milli\second}.

We explored various potentials and extensions of a \ac{SOME/IP} \ac{SD}-aware \ac{SDN} controller. 
Future work will focus on further optimizing service discovery, mobility, and reconfiguration mechanisms for improved robustness, QoS support, and security enhancements.


\bibliographystyle{IEEEtran}
\bibliography{bib/own,bibtex/simulation,bib/rfcs,bibtex/standards,bibtex/sdn,bibtex/soa,bibtex/security,bbl/bibliography}

\end{document}